\documentclass{article}

\usepackage{arxiv}

\usepackage[utf8]{inputenc} % allow utf-8 input
\usepackage[T1]{fontenc}    % use 8-bit T1 fonts
\usepackage{hyperref}       % hyperlinks
\usepackage{url}            % simple URL typesetting
\usepackage{booktabs}       % professional-quality tables
\usepackage{amsfonts}       % blackboard math symbols
\usepackage{nicefrac}       % compact symbols for 1/2, etc.
\usepackage{microtype}      % microtypography
\usepackage{lipsum}		% Can be removed after putting your text content
\usepackage{graphicx}
\usepackage{doi}
\title{Calibration method for complex permittivity measurements using s-SNOM combining multiple tapping harmonics}

\author{ 
    \href{https://orcid.org/0000-0002-5118-7368}{\includegraphics[scale=0.06]{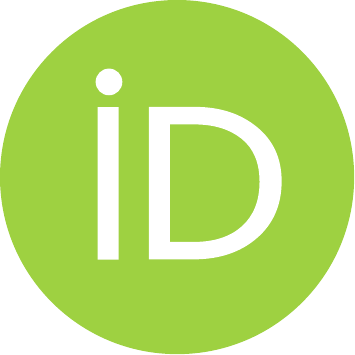}\hspace{1mm}Dario~Siebenkotten} \\
Physikalisch-Technische Bundesanstalt\\
	Abbestr. 2-12\\
	10587 Berlin, Germany\\
	\And
    \href{https://orcid.org/0000-0002-6575-6621}{\includegraphics[scale=0.06]{orcid.pdf}\hspace{1mm}Bernd~K\"astner}\thanks{Corresponding author: bernd.kaestner@ptb.de} \\
	Physikalisch-Technische Bundesanstalt\\
	Abbestr. 2-12\\
	10587 Berlin, Germany\\
	\And
    \href{https://orcid.org/0000-0003-0813-7514}{\includegraphics[scale=0.06]
    {orcid.pdf}\hspace{1mm}Arne~Hoehl}\\
	Physikalisch-Technische Bundesanstalt\\
	Abbestr. 2-12\\
	10587 Berlin, Germany\\
	\And
	\href{https://orcid.org/0000-0002-1596-6604}{\includegraphics[scale=0.06]{orcid.pdf}\hspace{1mm}Shuhei~Amakawa}\\ 
    Graduate School of Advanced Science and Engineering\\
	Hiroshima University\\
	Higashihiroshima, Japan\\
}

\renewcommand{\headeright}{Preprint}
\renewcommand{\undertitle}{Preprint}
\renewcommand{\shorttitle}{Calibration method for complex permittivity measurements using s-SNOM}

\begin{document}
\maketitle

\begin{abstract}
Scattering-type scanning near-field optical microscopy (s-SNOM) enables sub-diffraction spectroscopy,
featuring high sensitivity to small spatial permittivity variations of the sample surface. However, due to the 
near-field probe-sample interaction, the quantitative extraction of the complex permittivity leads to a computationally demanding inverse problem, requiring further approximation of the system to an invertible model.
Black-box calibration methods, similar to those applied to microwave vector network analysers, allow the extraction of the permittivity without detailed electromagnetic modelling of the probe-sample interaction.
These methods, however, are typically designed for stationary setups. In contrast, the distance between the sample and the probe tip of the s-SNOM is slowly modulated, which is required for the lock-in detection used to extract the near-field interaction buried in the far-field background. Here we propose an improved calibration method that explicitly takes probe tapping into account. We validate our method for an s-SNOM operating in a 
mid-infrared spectral range by applying it to measurements of silicon microstructures of different but well characterised doping.
\end{abstract}

% keywords can be removed

\section{Introduction}

Scattering-type scanning near-field optical microscopy (s-SNOM)~\cite{keilmann2004} enables sub-diffraction Fourier transform infrared nanospectroscopy (nano-FTIR)~\cite{huth2012,Hermann2013} by focussing electromagnetic radiation of a desired spectral range onto the metalised probe of an atomic force microscope (AFM). 
The probe can be regarded as an optical antenna~\cite{olmon2012},
locally coupling the incident radiation 
with the sample surface, with a spatial resolution determined by the probe-tip radius. The frequency-domain electric field phasor, 
$E_\mathrm{out}$, 
coming out of the s-SNOM system after being scattered by the probe, depends on the permittivity of the sample due to the near-field interaction between the probe tip and the sample. The interferometric nano-FTIR detection scheme allows phase-resolved detection of $E_\mathrm{out}=S E_\mathrm{in}$, where $E_\mathrm{in}$ is the field incident on the s-SNOM system, and $S$ is its complex-valued scattering coefficient~\cite{huth2012}, in which the complex permittivity of the sample is encoded.
The aim of quantitative s-SNOM measurements
(as opposed to qualitative imaging)
is to extract the spatial distribution of the complex permittivity of the 
sample under test.
This usually necessitates computationally expensive detailed electromagnetic modelling of the probe and the sample, which include assumptions about the shape of the probe and its optical properties~\cite{Cvitkovic2007, LightningRodModel, GeneralizedSpectralModelSNOM,McArdleSNOMSimulation}, and further approximation of the system to an invertible model~\cite{Govyadinov2014}.
In principle, alternative calibration methods based on black-box models
allow the extraction of the permittivity without detailed electromagnetic modelling of the probe-sample interaction. 
In this sense, the problem is comparable with that of scanning microwave microscopy (SMM), where 
permittivity measurement techniques in the GHz frequency range have been developed~\cite{Tanbakuchia2009b,Hoffmann2012b,Gramse2014,Horibe:mcosmmmsva}
using a vector network analyser (VNA)~\cite{Dunsmore:homcmwavt2}.
In their pioneering work, Guo~\emph{et al.}\ recently applied a similar calibration method to an s-SNOM at THz frequencies~\cite{Guo2021a}.  They subsequently extended it to multi-layer samples~\cite{Guo2023ML}. 
However, such black-box calibration methods are typically designed for stationary setups~\cite{Dunsmore:homcmwavt2}. 
In contrast, the distance between the probe tip and the sample of the s-SNOM is slowly modulated~\cite{HillenbrandDemodulation}, so that the information about the
near-field interaction, buried in the far-field background, can be extracted by high-order lock-in detection.  However, methods of dealing with the tapping of the probe within such black-box calibration approaches do not appear to have been reported so far. 
Here we propose an improved calibration method that explicitly takes 
the probe tapping into consideration. We validate our method for an s-SNOM operating in the mid-infrared spectral range by applying it to measurements of silicon %nanostructures 
microstructures of different but well characterised doping.

\section{Sketch of calibration and envelope-domain reconstruction}

\begin{figure}[tb]
\centering
\includegraphics[width=0.9\columnwidth]{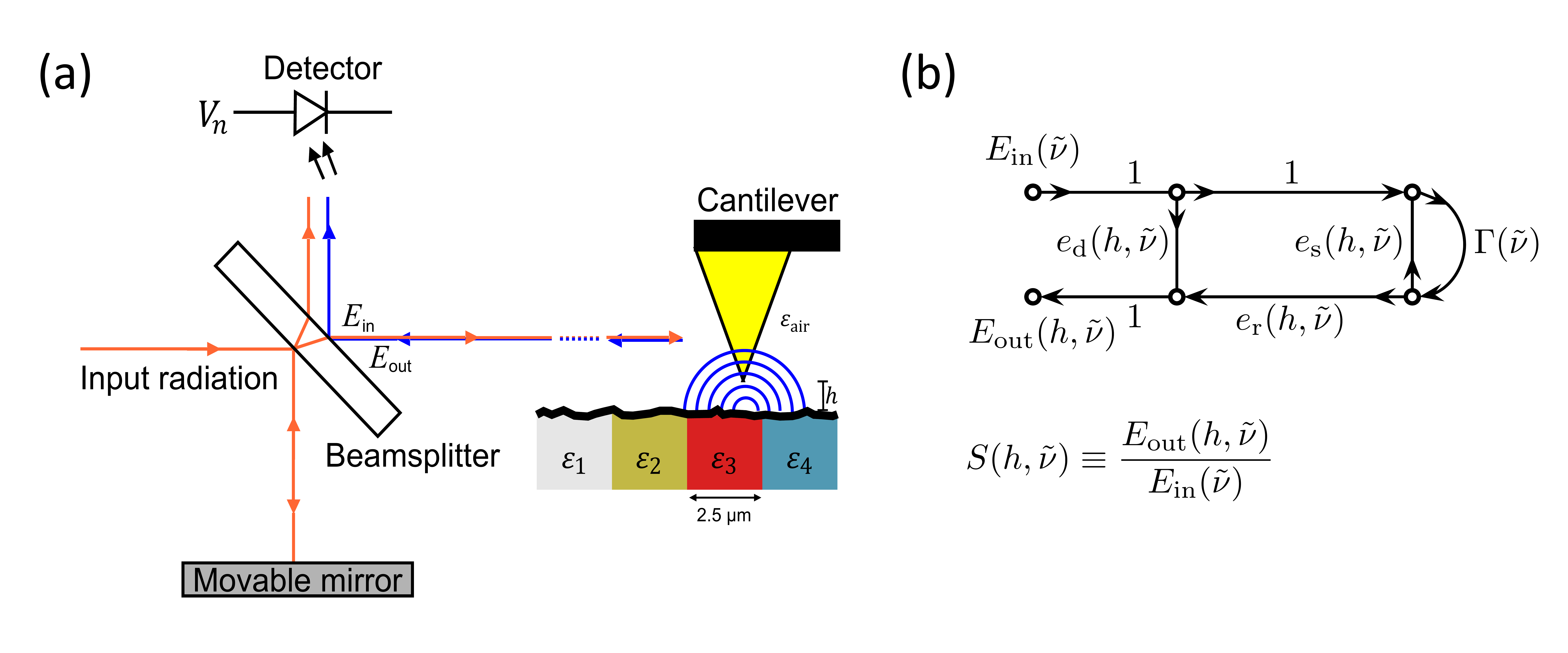}
\caption{
(a) Schematic optical beam paths as used in the broadband nano-FTIR measurements.
Infrared radiation from the storage ring is split into a reference beam, directed to a movable mirror, and a sample beam. The latter is focussed on the probe tip by optics (not shown) and scattered back, carrying 
information about the sample. The reference beam and the scattered sample beam then interfere at the detector.
(b)~A signal flow graph for the adapter that abstracts the far-field-to-near-field conversion via the probe.
$E_\mathrm{in}(\tilde{\nu})$ is the travelling wave incident on the s-SNOM system.  $E_\mathrm{out}(h,\tilde{\nu})$ is the outgoing (scattered) travelling wave. $S(h,\tilde{\nu})$ is the complex-valued scattering coefficient of the s-SNOM system.  $\Gamma(\tilde{\nu})$ is the reflection coefficient of the sample under test, assumed to be independent of the probe-tip height $h$. $e_\mathrm{d}(h,\tilde{\nu})$ represents sample-independent scattering of the incident wave $E_\mathrm{in}(\tilde{\nu})$ by the s-SNOM optics. $e_\mathrm{s}(h,\tilde{\nu})$ represents the reflection by the s-SNOM system of the near field reflected by the sample. 
$e_\mathrm{r}(h,\tilde{\nu})$ is the total transmission coefficient of the s-SNOM optics.
}
\label{fig:SetupAndSignalFlow}
\end{figure}

The nano-FTIR setup used for the validation is shown in Fig.~\ref{fig:SetupAndSignalFlow}(a). Broadband infrared synchrotron radiation, provided by the electron storage ring Metrology Light Source~\cite{gottwald2012, Hermann2013}, is used in an asymmetric Michelson interferometer, in which the probe above the sample is placed at the end of 
one arm, and a movable mirror 
at the end of the other. When the reference mirror is moved, the optical path difference 
between the two arms changes, and consequently so does the measured detector voltage.
This mechanism is used to interferometrically extract the detector voltage $V$
as a function of the frequency $\nu$, or equivalently the wavenumber $\tilde{\nu}\equiv\nu/c=1/\lambda$,
with $c$ being the speed of light in a vacuum, and $\lambda$ the wavelength.
This is in contrast with narrowband measurement systems, and complicates frequency/wavenumber-resolved measurements somewhat.  In a VNA-based SMM, for example, a swept monochromatic signal source is used.
%Further details about the components of our measurement setup can be found in Supplementary Material S1. 

Fig.~\ref{fig:SetupAndSignalFlow}(b) shows the signal flow graph~\cite{Mason:ftsposfg,Kuhn:ssfga} for the black-box ``far-field-to-near-field adapter'' model~\cite{Guo2021a}, in which 
%$e_\mathrm{d}(h,\nu)$, $e_\mathrm{h}(h,\nu)$, and $e_\mathrm{r}(h,\nu)$ 
$e_\mathrm{d}(h,\tilde{\nu})$, $e_\mathrm{h}(h,\tilde{\nu})$, and $e_\mathrm{r}(h,\tilde{\nu})$ are the 
adapter coefficients that depend, in general, 
on the distance $h$ between the probe tip and the sample surface and on the 
wavenumber $\tilde{\nu}$.  
$e_\mathrm{d}(h,\tilde{\nu})$, $e_\mathrm{h}(h,\tilde{\nu})$, and $e_\mathrm{r}(h,\tilde{\nu})$ are called the directivity, source match, and reflection tracking coefficients, respectively, in VNA calibration terminology~\cite{Dunsmore:homcmwavt2}, but these names may as well be forgotten in the s-SNOM context because they may have somewhat different meanings~\cite{Guo2021a}. 
From Fig.~\ref{fig:SetupAndSignalFlow}(b), we obtain, using the known graph reduction rules~\cite{Kuhn:ssfga}, the following formula describing the height- and 
wavenumber-dependent scattering coefficient $S(h,\tilde{\nu})$:
\begin{equation}
S(h,\tilde{\nu}) \equiv\frac{E_\mathrm{out}(h,\tilde{\nu})}{E_\mathrm{in}(\tilde{\nu})} = e_\mathrm{d}(h,\tilde{\nu}) + \frac{e_\mathrm{r}(h,\tilde{\nu}) \Gamma(\tilde{\nu})}{1- e_\mathrm{s}(h,\tilde{\nu}) \Gamma(\tilde{\nu})},
    \label{eq:errAdapt}
\end{equation}
where 
$\Gamma(\tilde{\nu})$ is the complex reflection coefficient of the sample.
$\Gamma(\tilde{\nu})$ is related to the complex permittivity $\varepsilon(\tilde{\nu})$ of the sample as follows~\cite{Hauer:qamfsinfmols}: %\cite{JacksonThirdEdition}
\begin{equation}
\Gamma(\tilde{\nu}) = \frac{\varepsilon(\tilde{\nu})-1}{\varepsilon(\tilde{\nu})+1}.
    \label{eq:Gamma:epsilon}
\end{equation}
%A derivation of Eq.~(\ref{eq:Gamma:epsilon}) is given in Supplementary Material S2. 
Note that $\Gamma(\tilde{\nu})$ is a sample property and therefore independent of $h$.
The sample reflection coefficient can be determined by using the following formula derived from Eq.~(\ref{eq:errAdapt}), if the values of the adapter coefficients
are known:
\begin{equation}
\Gamma(\tilde{\nu}) = \frac{S(h,\tilde{\nu}) - e_\mathrm{d}(h,\tilde{\nu})}{\left[S(h,\tilde{\nu}) - e_\mathrm{d}(h,\tilde{\nu})\right] e_\mathrm{s}(h,\tilde{\nu}) + e_\mathrm{r}(h,\tilde{\nu})}.
    \label{eq:GammaDet}
\end{equation}

For a given probe-tip height $h$, the adapter can be \emph{calibrated} or equivalently, the coefficients can be determined, by making three 
(or more)
calibration measurements; that is, by measuring the scattering coefficients, $S^{(m)}(h,\tilde{\nu})$ $(m=1,\ 2,\ 3)$, of three materials (or \emph{calibration standards}) with 
known and sufficiently different
reflection coefficients, $\Gamma^{(m)}(\tilde{\nu})$ $(m=1,\ 2,\ 3)$. 
Inserting these values in Eq.~(\ref{eq:errAdapt}) gives the simultaneous equations that can be numerically solved for 
$e_\mathrm{r}(h,\tilde{\nu})$, $e_\mathrm{d}(h,\tilde{\nu})$, and $e_\mathrm{s}(h,\tilde{\nu})$.

Although the complex scattering coefficient $S(h,\tilde{\nu})$ as defined in Eq.~(\ref{eq:errAdapt}) is a frequency-domain 
quantity, it can be considered to depend periodically on time because of the \emph{slow} modulation of the probe-tip height, $h(t)$, at an angular frequency $\Omega$, where $\Omega$ is orders of magnitude smaller than the mid-infrared frequency
of $E_\mathrm{in}(\tilde{\nu})$ and $E_\mathrm{out}(h,\tilde{\nu})$: $\Omega\ll 2\pi\nu$. Such a slowly time-varying component is called an \emph{envelope}~\cite{Verspecht:lsna}. The time-dependent envelope-domain (i.e., mixed time- and frequency-domain)~\cite{Verspecht:lsna} $S(h(t),\tilde{\nu})$ contains harmonic components at $n \Omega$ $(n = 0,\ 1,\ 2,\ \cdots)$ because the scattering depends nonlinearly on the height $h(t)$.  The slowly time-varying scattering coefficient can then be written as a Fourier series:
\begin{equation}
    S(h(t),\tilde{\nu}) = \sum_{n=-\infty}^{\infty} S_n(\tilde{\nu}) \exp\left({-\mathrm{i} n \Omega t}\right),
    \label{eq:sMod}
\end{equation}
where the complex-valued Fourier coefficients $S_n(\tilde{\nu})$ are independent of time.  The detector voltage phasors, $V_n(\tilde{\nu})$, obtained by $n$th-order lock-in detection (or demodulation) can be considered to be proportional to $S_n(\tilde{\nu})$, that is, $V_n(\tilde{\nu})\propto S_n(\tilde{\nu})$.  We will use this fact in the next section to establish the actual calibration equations to be used.
Note also that the adapter coefficients in Eq.~(\ref{eq:GammaDet}) also vary slowly and periodically in time \emph{in such a way that makes} $\Gamma(\tilde{\nu})$ \emph{time-invariant}. 

\begin{figure}[t]
\centering
\includegraphics[width=0.5\columnwidth]{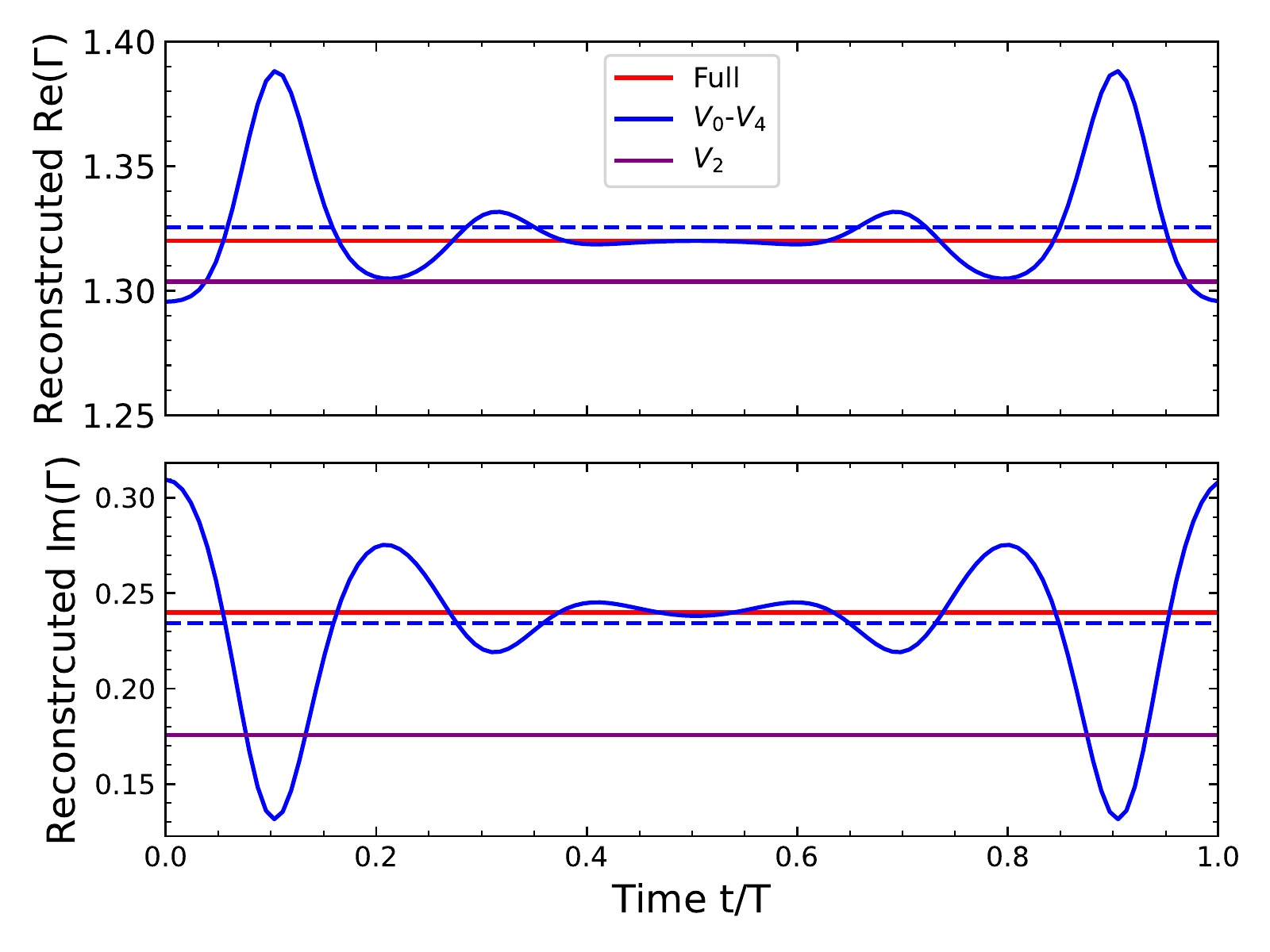}
 \caption{Shown are the reconstructed real and imaginary parts of $\Gamma$ versus time, plotted over one period $T\equiv 2\pi/\Omega$ for $\Gamma = 1.32 + 0.24\mathrm{i}$. Solid lines are the envelope-domain reconstructions, dashed lines the temporal averages.}
 \label{fig:Gamma_reconstruction}
\end{figure}

The importance of using height-dependent adapter coefficients for determining the sample reflection coefficient $\Gamma(\tilde{\nu})$ can be demonstrated by simulations using the finite dipole model~\cite{Cvitkovic2007}, which calculates $S(h,\tilde{\nu})$ quasi-statically for different heights. We assume here that the height measured from the surface is given by
\begin{equation}
\label{eq:h(t)}
    h(t) = \hat{h} \left[1+\cos\left( \Omega t \right)\right]
\end{equation}
over a probe tapping period $T\equiv 2\pi/\Omega$
at a wavenumber $\tilde{\nu}$, with $\hat{h}$ being the amplitude of the probe height modulation. 
The value of $\Gamma(\tilde{\nu})$ to be measured by the s-SNOM has been chosen to be non-trivial; that is, different from 
$\Gamma^{(m)}(\tilde{\nu})$ $(m=1,\ 2,\ 3)$ used for calibration. 
The result of using Eq.~(\ref{eq:sMod}) and time-dependent adapter coefficients for reconstructing the value of $\Gamma(\tilde{\nu})$ over one period $T$ is shown in Fig.~\ref{fig:Gamma_reconstruction} (red straight lines).
Although all quantities on the right-hand side of Eq.~(\ref{eq:GammaDet}) 
change with time, the resulting $\Gamma(\tilde{\nu})$ is constant and correct as it should be.

Also shown in Fig.~\ref{fig:Gamma_reconstruction} are the cases of including only the terms with $0\le n\le4$
and with $n=2$ for both calibration and reconstruction. The $\Gamma(\tilde{\nu})$ reconstructed from 
$0\le n\le4$ (blue curves) shows periodic deviations that grow towards the start and the end of the period. For this case, excellent agreement with the correct value is reached only around the centre of the period ($t=T/2$), corresponding to the minimal tip-sample distance. In contrast, the use of only $n=2$ (because of the assumed use of the second-order lock-in detection) leads to a constant deviation from the correct value 
(purple straight lines). This illustrates that the use of multiple tapping harmonics, rather than just a single one,  for the envelope-domain reconstruction is essential.

In the next section, we show that continuous-time (or fine-time-step) tracking of the adapter coefficients over a modulation period $T$ is, actually, not necessary and establish the calibration equations used.

\section{Calibration equations}

In order to establish a calibration procedure allowing for tip-height modulation, we need to consider the typical number of higher harmonics that can be measured with acceptable signal-to-noise ratio. Here we assume that measurements up to the 4th tapping harmonic are possible.  Although the result we saw in Fig.~\ref{fig:Gamma_reconstruction} might arouse some concern for using only $n\le 4$, it also suggests that the choice of right timing may give a good result.  Eq.~(\ref{eq:sMod}), together with Eq.~(\ref{eq:errAdapt}), then reduces to the following approximation:
\begin{equation}
    e_\mathrm{d}(t,\tilde{\nu}) + \frac{ e_\mathrm{r}(t,\tilde{\nu}) \Gamma(\tilde{\nu})}{1 - e_\mathrm{s}(t,\tilde{\nu}) \Gamma(\tilde{\nu})} %\right) 
\approx \sum_{n=-4}^{4} S_n(\tilde{\nu}) \exp\left({-\mathrm{i}n \Omega t}\right).
    \label{eq:calibTimedep}
\end{equation}
Note that we used a shorthand notation for the adapter coefficients in Eq.~(\ref{eq:calibTimedep}), ignoring the fact that they have different functional dependences on $t$ and $h(t)$ (cf.\ Eq.~(\ref{eq:errAdapt})). For air, $\Gamma(\tilde{\nu})$ is zero in the infrared spectral range. Therefore, according to Eq.~(\ref{eq:Gamma:epsilon}), $e_\mathrm{d}(t,\tilde{\nu})$ does not depend on $t$ or $\tilde{\nu}$. As the tapping amplitude $\hat{h}$ is kept small compared to the wavelength $\lambda$, no variation of the scattered signal due to tapping in air is expected. Thus, $S_n^{\mathrm{air}} = 0$ for $n>0$, and  $e_\mathrm{d}(t,\tilde{\nu})$ can be identified as
\begin{equation}
    e_\mathrm{d}\equiv S_0^\mathrm{air}.
    \label{eq:sAir}
\end{equation}

Since $S_0^\mathrm{air}$ is usually hard to measure directly, we assume that for a tapping amplitude $\hat{h}$ much larger than the probe tip radius, the electrical near-field has decayed sufficiently at time 
$t=t_\mathrm{max}$ 
($t=0$ in Eq.~(\ref{eq:h(t)}), for example), when the probe tip is at its maximum distance from the sample surface \cite{MooshammerFEM}, such that 
\begin{equation}
    S(t_\mathrm{max},\tilde{\nu}) = \sum_{n=-4}^4 S_n(\tilde{\nu}) \exp\left({-\mathrm{i}n \Omega t_\mathrm{max}}\right) \approx S_0^\mathrm{air}
    \label{eq:sAirApprox}
\end{equation}
holds. This leads to
\begin{equation}
    S_0(\tilde{\nu}) \approx S_0^\mathrm{air} - 2 \sum_{n = 1}^4 S_n(\tilde{\nu}) \cos \left(n \Omega t_\mathrm{max} \right),
\end{equation}
where we used $S_n(\tilde{\nu}) = S_{-n}(\tilde{\nu})$. Then, $S(t,\tilde{\nu})$, which also is a shorthand notation, can be written as
\begin{equation}
    S(t,\tilde{\nu}) \approx S_0^\mathrm{air} + 2 \sum_{n = 1}^4 \left[ \cos \left({n \Omega t}\right) - \cos \left({n \Omega t_\mathrm{max}}\right) \right] S_n(\tilde{\nu}).
    \label{eq:st}
\end{equation}
Defining 
\begin{equation}
    \hat{S}(t,\tilde{\nu}) \equiv 2 \sum_{n = 1}^4 \left[ \cos \left({n \Omega t}\right) - \cos \left({n \Omega t_\mathrm{max}}\right) \right] S_n(\tilde{\nu})
    \label{eq:hat[S](t):def}
\end{equation}
and combining equations (\ref{eq:calibTimedep}), (\ref{eq:sAir}), and (\ref{eq:st}), the term $S_0^\mathrm{air}$ drops out, and one obtains
\begin{equation}
       \frac{e_\mathrm{r}(t,\tilde{\nu}) \Gamma(\tilde{\nu})}{1 - e_\mathrm{s}(t,\tilde{\nu}) \Gamma(\tilde{\nu})} = \hat{S}(t,\tilde{\nu}).
\label{eq:S_0^air_dropped}
\end{equation}

In principle, this equation could be used for the adapter calibration, provided the scattering coefficient, $S(h,\tilde{\nu})$ in Eq.~(\ref{eq:errAdapt}),
or its transformed version, $\hat{S}(t,\tilde{\nu})$ 
in Eq.~(\ref{eq:hat[S](t):def}), 
is measurable.  However, unlike in a VNA, %$S(h,\tilde{\nu})$ 
the scattering coefficient is not measured in our interferometric measurement setup. What are actually measured instead are the detector voltage phasors, $V_n(\tilde{\nu})$, which are proportional to the Fourier components, $S_n(\tilde{\nu})$, in Eq.~(\ref{eq:sMod}).
Taking the ratio of Eq.~(\ref{eq:S_0^air_dropped}) at two different times, $t_0$ and $t_1$, during a modulation cycle eliminates the need to determine the proportionality constant. 
\begin{equation}
\frac{\left[1 - e_\mathrm{s}(t_1,\tilde{\nu}) \Gamma(\tilde{\nu})\right] e_\mathrm{r}(t_0,\tilde{\nu}) }{\left[1 - e_\mathrm{s}(t_0,\tilde{\nu}) \Gamma(\tilde{\nu})\right]e_\mathrm{r}(t_1,\tilde{\nu})}  = \frac{\hat{S}(t_{0},\tilde{\nu})}{\hat{S}(t_{1},\tilde{\nu})} = \frac{\hat{V}(t_{0},\tilde{\nu})}{\hat{V}(t_{1},\tilde{\nu})},
\label{eq:ratio_t0_t1}
\end{equation}
where, just as in Eq.~(\ref{eq:hat[S](t):def}),
\begin{equation}
    \hat{V}(t,\tilde{\nu}) \equiv 2 \sum_{n = 1}^4 \left[ \cos \left({n \Omega t}\right) - \cos \left({n \Omega t_\mathrm{max}}\right) \right] V_n(\tilde{\nu}).
    \label{eq:hat[V](t):def}
\end{equation}
There could actually be slight systematic deviations between calibration measurements. Such systematic deviations could result from slow temporal fluctuations of the light source and from unwanted reflections from the sample surface~\cite{Cvitkovic2007, RaschkeReflection, mester2021}. Taking a ratio as in Eq.~(\ref{eq:ratio_t0_t1}), using two time instants of a single measurement, has the added benefit of mitigating the adverse effects of such systematic deviations.

Now $e_\mathrm{r}(t_0,\tilde{\nu})$ and $e_\mathrm{r}(t_1,\tilde{\nu})$ in Eq.~(\ref{eq:ratio_t0_t1}) can be eliminated by using Eqs.~(\ref{eq:S_0^air_dropped}) and (\ref{eq:hat[V](t):def}) for, without loss of generality, the third calibration measurement. Thus, we obtain 
\begin{equation}
\frac{\left[1 - e_\mathrm{s}(t_1,\tilde{\nu}) \Gamma(\tilde{\nu})\right] \left[1 - e_\mathrm{s}(t_0,\tilde{\nu}) \Gamma^{(3)}(\tilde{\nu})\right]}{ \left[1 - e_\mathrm{s}(t_0,\tilde{\nu}) \Gamma(\tilde{\nu})\right]\left[1 - e_\mathrm{s}(t_1,\tilde{\nu}) \Gamma^{(3)}(\tilde{\nu})\right]} 
= \frac{\hat{V}(t_{0},\tilde{\nu})\hat{V}^{(3)}(t_{1},\tilde{\nu})}{\hat{V}(t_{1},\tilde{\nu})\hat{V}^{(3)}(t_{0},\tilde{\nu})},
           \label{eq:Model}
\end{equation}
where 
$\hat{V}^{(3)}(t_{0},\tilde{\nu})$ and $\hat{V}^{(3)}(t_{1},\tilde{\nu})$ 
are the values of 
Eq.~(\ref{eq:hat[V](t):def}) from the third calibration measurement.
The remaining unknowns to be determined in Eq.~(\ref{eq:Model}) are $e_\mathrm{s}(t_0,\tilde{\nu})$ and $e_\mathrm{s}(t_1,\tilde{\nu})$. A set of simultaneous equations for determining these can be obtained by inserting the results of the first and the second calibration measurements in Eq.~(\ref{eq:Model}) as follows:
\begin{equation}
\left\{\begin{array}{l}
\displaystyle
\frac{\left[1 - e_\mathrm{s}(t_1,\tilde{\nu}) \Gamma^{(1)}(\tilde{\nu})\right] \left[1 - e_\mathrm{s}(t_0,\tilde{\nu}) \Gamma^{(3)}(\tilde{\nu})\right]}{ \left[1 - e_\mathrm{s}(t_0,\tilde{\nu}) \Gamma^{(1)}(\tilde{\nu})\right]\left[1 - e_\mathrm{s}(t_1,\tilde{\nu}) \Gamma^{(3)}(\tilde{\nu})\right]}  = \frac{\hat{V}^{(1)}(t_{0},\tilde{\nu})\hat{V}^{(3)}(t_{1},\tilde{\nu})}{\hat{V}^{(1)}(t_{1},\tilde{\nu})\hat{V}^{(3)}(t_{0},\tilde{\nu})}\\
\\
\displaystyle
\frac{\left[1 - e_\mathrm{s}(t_1,\tilde{\nu}) \Gamma^{(2)}(\tilde{\nu})\right] \left[1 - e_\mathrm{s}(t_0,\tilde{\nu}) \Gamma^{(3)}(\tilde{\nu})\right]}{ \left[1 - e_\mathrm{s}(t_0,\tilde{\nu}) \Gamma^{(2)}(\tilde{\nu})\right]\left[1 - e_\mathrm{s}(t_1,\tilde{\nu}) \Gamma^{(3)}(\tilde{\nu})\right]}  = \frac{\hat{V}^{(2)}(t_{0},\tilde{\nu})\hat{V}^{(3)}(t_{1},\tilde{\nu})}{\hat{V}^{(2)}(t_{1},\tilde{\nu})\hat{V}^{(3)}(t_{0},\tilde{\nu})}
\end{array}\right.
\label{eq:Model2}
\end{equation}

%The solutions to these equations are shown in Supplementary Material S3.

\section{Results and discussion}

In the following, Eq.~(\ref{eq:Model2}) serves as the set of calibration equations.  We illustrate their use with the nano-FTIR setup introduced above (Fig.~\ref{fig:SetupAndSignalFlow}(a)). 

To validate the calibration method, we apply it to a commercially available silicon sample, featuring several 2.5-\textmu m wide n-type doped stripes of different doping levels, manufactured and characterised via secondary ion mass spectroscopy by Infineon Technologies~\cite{SCHWEINBOCK20142070,Brinciotti2015,RitchieInfineonSNOM}. Four differently doped stripes are used for the validation.  These will hereafter be referred to as stripes 1 to 4, in the order of increasing doping. The doping concentration of stripe~1 is significantly lower than those of the other three stripes, 
making it indistinguishable from undoped silicon in the infrared spectral range. Three stripes (1, 2, and 4) are used as calibration standards, while stripe~3 is treated as the unknown sample under test. The permittivity $\varepsilon (\tilde{\nu})$ of the doped Si is assumed to follow the Drude model for free electrons~\cite{ashcroft1976solid,Bohren:aasolbsp}:
\begin{equation}
    \varepsilon (\tilde{\nu}) = \varepsilon_{\infty}\left(1-\frac{\tilde{\nu}_\mathrm{p}^2}{\tilde{\nu}^2+ \mathrm{i}\gamma \tilde{\nu}}\right),
    \label{eq:Drude}
\end{equation}
where $\varepsilon_{\infty} \approx 11.7$ is the high-frequency relative permittivity of Si~\cite{SiElectricalCharacteristics}, $\gamma$ is the damping rate, and 
\begin{equation}
    \tilde{\nu}_\mathrm{p} = \frac{1}{2\pi c}\sqrt{\frac{N e^2}{\varepsilon_0\varepsilon_{\infty}m^*}},
    \label{eq:PlasmaFreq}
\end{equation}
is the wavenumber corresponding to the plasma frequency, 
with $e$ being the elementary charge, $\varepsilon_0$ the vacuum permittivity, $N$ the free electron density, and $m^*$ the effective electron mass, which approximately equals $0.26\,m_\mathrm{e}$ 
in Si~\cite{SiElectricalCharacteristics}, where $m_\mathrm{e}$ is the electron mass. The damping rate, which is inversely proportional to the electron mobility~\cite{ashcroft1976solid}, was determined by using the relation described by Arora~\emph{et al.}~\cite{Arora1982} for each of the calibration stripes.

\begin{figure*}[tb]
\centering
\includegraphics[width=0.95\columnwidth]{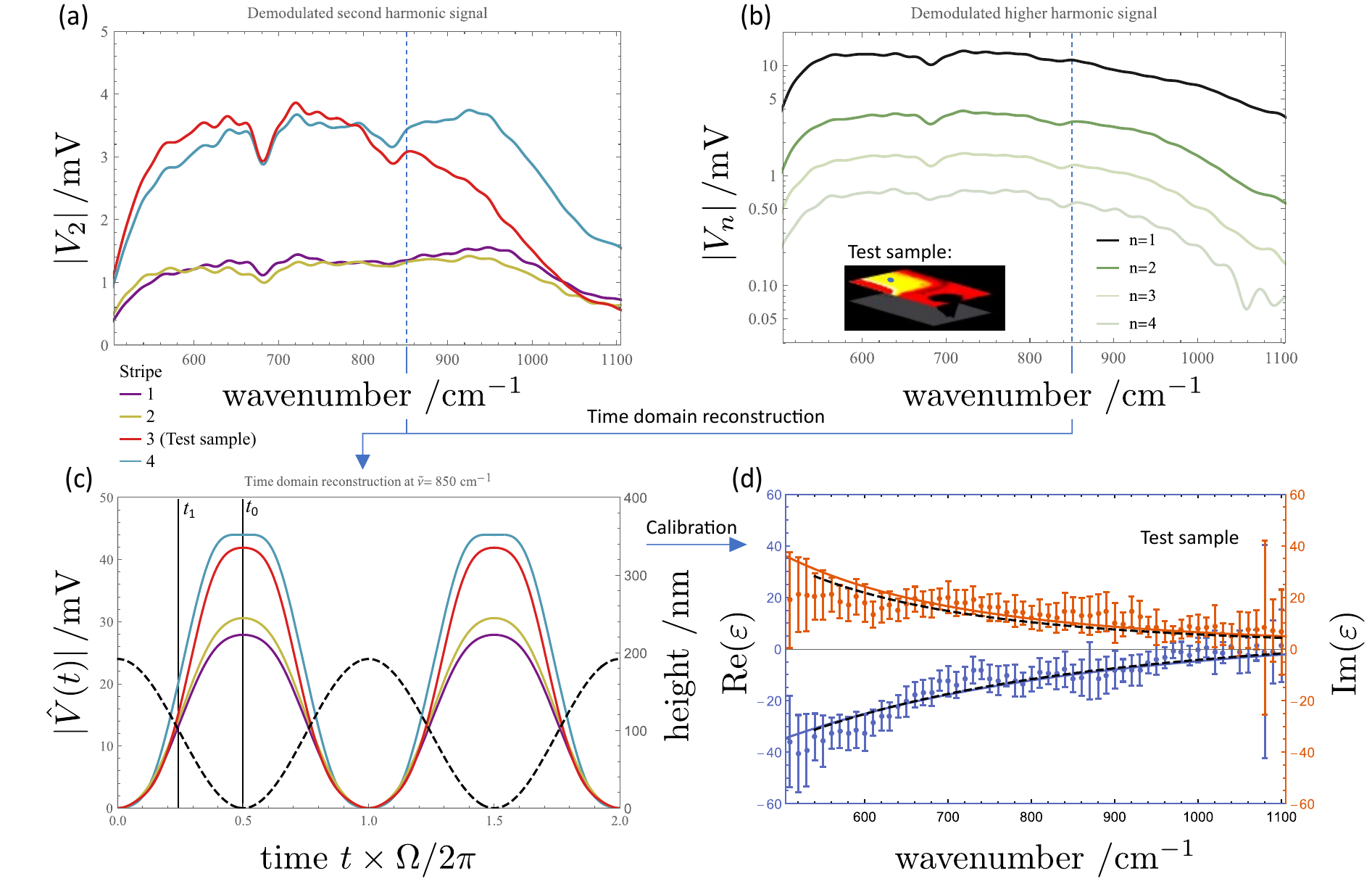}
\caption{\label{fig:mainFig}
(a)~Magnitude spectra of the second tapping harmonic of the detector voltage $|V_2(\tilde{\nu})|$ of all four stripes of different doping. (b)~Different tapping harmonics of the test sample (stripe 3). Inset: combined topography and optical image ($3\times 5$\,\textmu m$^2$) at a fixed mirror position of the test stripe. The probe position for the measurement is indicated by the blue dot. (c)~Envelope-domain reconstruction of the signal amplitude over two tip oscillation periods at a fixed wavenumber of 850\,cm$^{-1}$ using the 1st to 4th harmonic. The dashed black line shows the distance variation between the probe tip and the sample. The vertical black lines mark the points in time, $t_0$ and $t_1$, used for the calibration. (d)~The real and imaginary parts of the permittivity $\varepsilon(\tilde{\nu})$ of the test sample obtained after the calibration using the other three stripes, together with vertical error bars. Solid curves show expected values according to the Drude model Eq.~(\ref{eq:Drude}) and manufacturer specifications (Table~\ref{table:results}). Dashed curves show the best fit of the Drude permittivity Eq.~(\ref{eq:Drude}) using the electron density $N$ and the damping rate $\gamma$ as the fitting parameters.}
\end{figure*}

Fig.~\ref{fig:mainFig}(a) shows the magnitude of the measured second tapping harmonic component 
$V_2(\tilde{\nu})$ in Eq.~(\ref{eq:hat[V](t):def}) versus the wavenumber $\tilde{\nu}$ for all four stripes.
Fig.~\ref{fig:mainFig}(b) shows $|V_n(\tilde{\nu})|$ $(n=1,\ 2,\ 3,\ 4)$ of stripe~3, which serves as the test sample. For each spectrum, ten measurements were recorded consecutively, and the resulting spectra averaged.  These averaged spectra are shown in Figs.~\ref{fig:mainFig}(a) and (b).
The topography combined with an optical image at a fixed interferometer position of the test stripe 3 is presented in the inset of (b). All spectral measurements were performed at the centre of each stripe, as indicated by the blue dot for the test stripe. 
Using all four tapping harmonics, the near-field signal $\hat{V}(t,\tilde{\nu})$ can be reconstructed in the envelope-domain, and its magnitude is shown in Fig.~\ref{fig:mainFig}(c) over two probe oscillation periods, $2T$, for a fixed wavenumber of $850$\,cm$^{-1}$. It changes synchronously with the probe oscillation (black dashed line), assumed here to be given by Eq.~(\ref{eq:h(t)}), with the minimal tip-sample distance corresponding to the maximum near-field intensity. The expected rapid increase in $|\hat{V}(t,\tilde{\nu})|$ is seen at intermediate distances. However, the near-field signal intensity increases less strongly for smaller distances than expected for the corresponding exponentially growing near-field interaction. This might indicate that the actual tapping is not fully sinusoidal as in Eq.~(\ref{eq:h(t)}). Note, however, that we did not use Eq.~(\ref{eq:h(t)}) in our formulation. Thus, our calibration approach is not affected by the possibility of tapping being non-sinusoidal. From the envelope-domain plot, the two points in time, $t_0$ and $t_1$, needed for the calibration, can be chosen. Here, $t_0$ is chosen to be $t_0=T/2$, and $t_1$ is chosen such that the probe is in downwards motion and the scattering amplitudes still differ significantly among the different stripes (1, 2, and 4); specifically, $t_1 = T/4$. The real and imaginary parts of the measured complex permittivity of the test sample are plotted with error bars in Fig.~\ref{fig:mainFig}(d), together with the nominal permittivity (solid curves) estimated from manufacturer specifications (Table~\ref{table:results}) and the Drude model. To validate the accuracy of the calibration, the Drude model Eq.~(\ref{eq:Drude}) is fitted to the measured permittivity between 540 and 1100\,cm$^{-1}$, with $N$ and $\gamma$ treated as fitting parameters. The best fit is plotted as the dashed black lines in Fig.~\ref{fig:mainFig}(d). 

The uncertainty of each point in the complex permittivity spectrum in Fig.~\ref{fig:mainFig}(d) was estimated based on the uncertainty of the nominal electron density and measurement uncertainties. An uncertainty of 5\% with a uniform distribution was assumed centred around the nominal electron densities of the calibration standards. The uncertainty of each harmonic of the measured data, $V_n(\tilde{\nu})$, was assumed to be normally distributed and determined from the variance of the ten spectra. These uncertainties are propagated to the complex permittivity spectrum by calculating the standard deviation of 1000 repetitions of the calibration with the input spectra and electron densities varied randomly within these limits. To each of these repeated calibrations the Drude model fit was performed individually, and the standard deviations of the extracted electron density and damping rate were calculated from the resulting variation. The software \emph{Wolfram Mathematica} was used for the calibration and uncertainty propagation~\cite{Mathematica}.

For further validation of the model, the other two highly doped stripes (stripes 2 and 4) were also individually analysed as unknown test samples, with the respective other three stripes used as calibration standards. The electron densities and damping rates extracted by fitting the resulting calibrated permittivities compared to their nominal values are presented in Table~\ref{table:results} for stripes 2, 3, and 4 as test samples. For all three stripes, the fitted electron densities and damping rates show very good agreement with the nominal values. It is particularly noteworthy that the damping rate was fitted as an independent parameter, whereas the nominal damping rate was calculated from the nominal electron density. Nevertheless, still very good agreement is achieved. This shows our method is capable of determining the local charge carrier density and mobility simultaneously. The extraction of the permittivity via fitting also reduces the uncertainty in the determined parameters significantly compared to the relatively large uncertainty in the individual permittivity values in Fig.~\ref{fig:mainFig}(d). 
%(cf. also Supplementary Material S4). 
It is apparent that for the highest analysed doping (stripe 4), the uncertainty of the damping rate is significantly larger than those for the other two stripes. This can likely be explained by the nearly metallic behaviour of that stripe in the measured wavenumber regime, where large deviations in permittivity only lead to very slight variations in the scattered field. 

\begin{table}[t]
\centering
\caption{Results of the Drude model fits to the calibrated permittivities for three different doping densities and comparison to their nominal values. For each stripe, the other two and the lowly doped stripe were used as calibration standards.}
\begin{tabular}{cccccc}
Stripe&$N_{\mathrm{nom}}$ (cm$^{-3}$)&$N_{\mathrm{meas}}$ (cm$^{-3}$)&$\gamma_{\mathrm{nom}}$ (cm$^{-1}$)&$\gamma_{\mathrm{meas}}$ (cm$^{-1}$)&\\
\hline
2&$1.2\times 10^{19}$&$(1.3\pm 0.1)\times 10^{19}$&$350$&$400\pm20$\rule{0pt}{2.6ex}\\
3&$5.5\times 10^{19}$&$(5.2\pm 0.2)\times 10^{19}$&$390$&$350\pm20$\\
4&$1.5\times 10^{20}$&$(1.0\pm 0.1)\times 10^{20}$&$400$&$380\pm60$\\
\end{tabular}
\label{table:results}
\end{table}

\section{Conclusion and outlook}

In conclusion, we developed a calibration method for nanoscale resolution permittivity measurement using s-SNOM, extending stationary black-box models used in scanning microwave microscopy and the work of Guo \emph{et al.}\ for s-SNOM in the THz regime~\cite{Guo2021a}. Our method takes probe tapping into account in extracting the time-invariant sample permittivity. To do so, multiple tapping harmonics of the measured detector voltage are used to take the slow temporal variation of the probe height into consideration. A decisive advantage of this calibration method, compared to conventional methods, is that no detailed knowledge or computationally expensive electromagnetic modelling of the probe is required. We validated the method by measuring Si microstructures of different doping levels. We extracted their respective electron densities and damping rates via fitting of the Drude model to the measured permittivities in the infrared spectral range. The extracted parameters showed very good agreement with the nominal values. 
While the uncertainties of the recovered permittivities are large in our example (Fig.~\ref{fig:mainFig}(d)), the uncertainties of the extracted electron density and damping rate are significantly smaller (Table~\ref{table:results}). In many applications, tunable lasers operating at a single wavelength can be used instead of broadband sources, which should significantly reduce the permittivity uncertainty. Further improvements in calibration accuracy are feasible by using calibration standards that are better characterised, thus known with smaller permittivity uncertainty.
Our model extracts the sample permittivity directly, so samples are not limited to those that exhibit Drude-like behaviour. Combined with the nanoscale resolution s-SNOM offers, the proposed method promises both sensitive and quantitative characterisation of electronic nanostructures and quantum devices.

\section{Acknowledgement}
We acknowledge fruitful discussions with Georg Gramse, Johannes Hoffmann, and Manuel Marschall.

\section{Funding}
The authors D.S., B.K. and A.H. have received funding from the project ``20IND12 ELENA'' within the EMPIR programme co-financed by the Participating States and from the European Union's Horizon 2020 research and innovation programme. B.K.\ and S.A.\ also received funding from the Japan Society for the Promotion of Science (Fellowship ID: S19133). S.A.\ was supported in part by JSPS.KAKENHI (22H00217) and MEXT Initiative to Establish Next-generation Novel Integrated Circuits Centers (X-NICS) Grant Number JPJ011438.

%\printbibliography  

\bibliographystyle{unsrt}

\bibliography{references}  %%% Uncomment this line and comment out the ``thebibliography'' section below to use the external .bib file (using bibtex) .

\begin{thebibliography}{10}

\bibitem{keilmann2004}
F.~Keilmann and R.~Hillenbrand.
\newblock Near-field microscopy by elastic light scattering from a tip.
\newblock {\em Phil. Trans. R. Soc. Lond. A}, 362:787--805, 2004.

\bibitem{huth2012}
Florian Huth, Alexander Govyadinov, Sergiu Amarie, Wiwat Nuansing, Fritz
  Keilmann, and Rainer Hillenbrand.
\newblock Nano-{FTIR} absorption spectroscopy of molecular fingerprints at 20
  nm spatial resolution.
\newblock {\em Nano Letters}, 12:3973--8, 8 2012.

\bibitem{Hermann2013}
Peter Hermann, Arne Hoehl, Piotr Patoka, Florian Huth, Eckart Rühl, and
  Gerhard Ulm.
\newblock Near-field imaging and nano-{Fourier}-transform infrared spectroscopy
  using broadband synchrotron radiation.
\newblock {\em Optics Express}, 21:2913, 2 2013.

\bibitem{olmon2012}
R~L Olmon and M~B Raschke.
\newblock Antenna–load interactions at optical frequencies: impedance
  matching to quantum systems.
\newblock {\em Nanotechnology}, 23:444001, 2012.

\bibitem{Cvitkovic2007}
A.~Cvitkovic, N.~Ocelic, and R.~Hillenbrand.
\newblock Analytical model for quantitative prediction of material contrasts in
  scattering-type near-field optical microscopy.
\newblock {\em Optics Express}, 15:8550, 2007.

\bibitem{LightningRodModel}
Alexander~S. McLeod, P.~Kelly, M.~D. Goldflam, Z.~Gainsforth, A.~J. Westphal,
  Gerardo Dominguez, Mark~H. Thiemens, Michael~M. Fogler, and D.~N. Basov.
\newblock Model for quantitative tip-enhanced spectroscopy and the extraction
  of nanoscale-resolved optical constants.
\newblock {\em Phys. Rev. B}, 90:085136, Aug 2014.

\bibitem{GeneralizedSpectralModelSNOM}
B.-Y. Jiang, L.~M. Zhang, A.~H. Castro~Neto, D.~N. Basov, and M.~M. Fogler.
\newblock Generalized spectral method for near-field optical microscopy.
\newblock {\em Journal of Applied Physics}, 119(5):054305, 2016.

\bibitem{McArdleSNOMSimulation}
P.~McArdle, D.~J. Lahneman, Amlan Biswas, F.~Keilmann, and M.~M. Qazilbash.
\newblock Near-field infrared nanospectroscopy of surface phonon-polariton
  resonances.
\newblock {\em Phys. Rev. Res.}, 2:023272, Jun 2020.

\bibitem{Govyadinov2014}
Alexander~A. Govyadinov, Stefan Mastel, Federico Golmar, Andrey Chuvilin,
  P.~Scott Carney, and Rainer Hillenbrand.
\newblock {Recovery of Permittivity and Depth from Near-Field Data as a Step
  Towards Optical Nanotomography.}
\newblock {\em ACS Nano}, 8(7):6911, jun 2014.

\bibitem{Tanbakuchia2009b}
Hassan Tanbakuchi, Matt Richter, Ferry Kienberger, and Hans-Peter Huber.
\newblock Nanoscale materials and device characterization via a scanning
  microwave microscope.
\newblock In {\em 2009 IEEE International Conference on Microwaves,
  Communications, Antennas and Electronics Systems}, 2009.

\bibitem{Hoffmann2012b}
Johannes Hoffmann, Michael Wollensack, Markus Zeier, Jens Niegemann, Hans-Peter
  Huber, and Ferry Kienberger.
\newblock A calibration algorithm for nearfield scanning microwave microscopes.
\newblock In {\em 2012 12th IEEE International Conference on Nanotechnology
  (IEEE-NANO)}, 2012.

\bibitem{Gramse2014}
G.~Gramse, M.~Kasper, L.~Fumagalli, G.~Gomila, P.~Hinterdorfer, and
  F.~Kienberger.
\newblock Calibrated complex impedance and permittivity measurements with
  scanning microwave microscopy.
\newblock {\em Nanotechnology}, 25, 4 2014.

\bibitem{Horibe:mcosmmmsva}
Masahiro Horibe, Seitaro Kon, and Iku Hirano.
\newblock Measurement capability of scanning microwave microscopy: Measurement
  sensitivity versus accuracy.
\newblock {\em IEEE Transactions on Instrumentation and Measurement},
  68(6):1774--1780, June 2019.

\bibitem{Dunsmore:homcmwavt2}
Joel~P. Dunsmore.
\newblock {\em Handbook of Microwave Component Measurements with Advanced {VNA}
  Techniques}.
\newblock Wiley, Hoboken, NJ, USA, second edition, 2020.

\bibitem{Guo2021a}
Xiao Guo, Karl Bertling, and Aleksandar~D. Rakić.
\newblock Optical constants from scattering-type scanning near-field optical
  microscope.
\newblock {\em Applied Physics Letters}, 118, 2021.

\bibitem{Guo2023ML}
Xiao Guo, Xin He, Zachary Degnan, Chun-Ching Chiu, Bogdan~C. Donose, Karl
  Bertling, Arkady Fedorov, Aleksandar~D. Rakić, and Peter Jacobson.
\newblock Terahertz nanospectroscopy of plasmon polaritons for the evaluation
  of doping in quantum devices.
\newblock {\em Nanophotonics}, 2023.

\bibitem{HillenbrandDemodulation}
Rainer Hillenbrand, B~Knoll, and Fritz Keilmann.
\newblock Pure optical contrast in scattering-type scanning near-field
  microscopy.
\newblock {\em Journal of Microscopy}, 202:77--83, 05 2001.

\bibitem{gottwald2012}
Alexander Gottwald, Roman Klein, Ralph Müller, Mathias Richter, Frank Scholze,
  Reiner Thornagel, and Gerhard Ulm.
\newblock Current capabilities at the metrology light source.
\newblock {\em Metrologia}, 49(2):S146, mar 2012.

\bibitem{Mason:ftsposfg}
Samuel~J. Mason.
\newblock Feedback theory---some properties of signal flow graphs.
\newblock {\em Proceedings of the IRE}, 41(9):1144--1156, September 1953.

\bibitem{Kuhn:ssfga}
Nicholas Kuhn.
\newblock Simplified signal flow graph analysis.
\newblock {\em Microwave Journal}, 6(11):59--66, November 1963.

\bibitem{Hauer:qamfsinfmols}
Benedikt Hauer, Andreas~P. Engelhardt, and Thomas Taubner.
\newblock Quasi-analytical model for scattering infrared near-field microscopy
  on layered systems.
\newblock {\em Optics Express}, 20(12):13173--13188, 2012.

\bibitem{Verspecht:lsna}
Jan Verspecht.
\newblock Large-signal network analysis.
\newblock {\em IEEE Microwave Mag.}, 6(4):82--92, December 2005.

\bibitem{MooshammerFEM}
Fabian Mooshammer, Markus~A. Huber, Fabian Sandner, Markus Plankl, Martin
  Zizlsperger, and Rupert Huber.
\newblock Quantifying nanoscale electromagnetic fields in near-field microscopy
  by fourier demodulation analysis.
\newblock {\em ACS Photonics}, 7(2):344--351, 2020.

\bibitem{RaschkeReflection}
Markus~B. Raschke and Christoph Lienau.
\newblock {Apertureless near-field optical microscopy: Tip–sample coupling in
  elastic light scattering}.
\newblock {\em Applied Physics Letters}, 83(24):5089--5091, 12 2003.

\bibitem{mester2021}
Lars Mester, Alexander~A Govyadinov, and Rainer Hillenbrand.
\newblock High-fidelity nano-{FTIR} spectroscopy by on-pixel normalization of
  signal harmonics.
\newblock {\em Nanophotonics}, 12 2021.

\bibitem{SCHWEINBOCK20142070}
T.~Schweinböck and S.~Hommel.
\newblock Quantitative scanning microwave microscopy: A calibration flow.
\newblock {\em Microelectronics Reliability}, 54(9):2070--2074, 2014.
\newblock SI: ESREF 2014.

\bibitem{Brinciotti2015}
Enrico Brinciotti, Georg Gramse, Soeren Hommel, Thomas Schweinboeck, Andreas
  Altes, Matthias~A. Fenner, Juergen Smoliner, Manuel Kasper, Giorgio Badino,
  Silviu-Sorin Tuca, and Ferry Kienberger.
\newblock Probing resistivity and doping concentration of semiconductors at the
  nanoscale using scanning microwave microscopy.
\newblock {\em Nanoscale}, 7:14715--14722, 2015.

\bibitem{RitchieInfineonSNOM}
Earl~T. Ritchie, Clayton~B. Casper, Taehyun~A. Lee, and Joanna~M. Atkin.
\newblock Quantitative local conductivity imaging of semiconductors using
  near-field optical microscopy.
\newblock {\em The Journal of Physical Chemistry C}, 126(9):4515--4521, 2022.

\bibitem{ashcroft1976solid}
N.W. Ashcroft and N.D. Mermin.
\newblock {\em Solid State Physics}.
\newblock HRW international editions. Holt, Rinehart and Winston, 1976.

\bibitem{Bohren:aasolbsp}
Craig~F. Bohren and Donald~R. Huffman.
\newblock {\em Absorption and Scattering of Light by Small Particles}.
\newblock Wiley, New York, NY, USA, 1983.

\bibitem{SiElectricalCharacteristics}
P.~Valizadeh.
\newblock {\em Field Effect Transistors, A Comprehensive Overview: From Basic
  Concepts to Novel Technologies}.
\newblock John Wiley and Sons Ltd, 03 2016.

\bibitem{Arora1982}
N.D. Arora, J.R. Hauser, and D.J. Roulston.
\newblock Electron and hole mobilities in silicon as a function of
  concentration and temperature.
\newblock {\em IEEE Transactions on Electron Devices}, 29(2):292--295, 1982.

\bibitem{Mathematica}
{Wolfram Research{,} Inc.}
\newblock Mathematica, {V}ersion 11.0.
\newblock Champaign, IL, 2021.

\end{thebibliography}





% \end{thebibliography}

\end{document}